\documentclass[journal,twocolumn,10pt]{IEEEtran}
\usepackage{multirow}

\usepackage{cite}
\usepackage{amsmath,amssymb,amsfonts}
\usepackage{algorithmic}
\usepackage{graphicx}
\usepackage{textcomp}
\usepackage{xcolor,colortbl,soul}
\usepackage{url}
\usepackage{array}
\usepackage{pgfplotstable}
\newtheorem{defn}{Definition}
\def\BibTeX{{\rm B\kern-.05em{\sc i\kern-.025em b}\kern-.08em
		T\kern-.1667em\lower.7ex\hbox{E}\kern-.125emX}}

\definecolor{Gray}{gray}{0.85}

\newcolumntype{g}{>{\columncolor{Gray}}c}

\def\cca#1{%
	\pgfmathsetmacro\calc{#1*100-0}%
	\edef\clrmacro{\noexpand\cellcolor{green!\calc}}%
	\clrmacro%
	\ifdim \calc pt>50pt\color{black}\fi{#1}%
}

\begin{document}


	\title{Machine Reasoning to Assess Pandemics Risks: Case of USS Theodore Roosevelt}
	\author{\IEEEauthorblockN{ 
			{Kenneth Lai and Svetlana N. Yanushkevich}
		}\\
		\IEEEauthorblockA{
			\textit{Biometric Technologies Laboratory, ECE Department,} \textit{University of Calgary, Canada} \\
			http://www.ucalgary.ca/btlab, \{kelai,syanshk\}@ucalgary.ca}
	}

	\markboth{IEEE Transactions on Emerging Topics in Computational Intelligence, ~Vol.~ , No.~2020}
	{...\MakeLowercase{\textit{Lai et al.}}:
		...}
	
	\maketitle

	\thispagestyle{empty}
	
	\begin{abstract}
		Assessment of risks of pandemics to communities and workplaces requires an intelligent decision support system (DSS). The core of such DSS must be based on machine reasoning techniques such as inference and shall be capable of estimating risks and biases in decision making. In this paper, we use a causal network to make Bayesian inference on COVID-19 data, in particular, assess risks such as infection rate and other precaution indicators. Unlike other statistical models, a Bayesian causal network combines various sources of data through joint distribution, and better reflects the uncertainty of the available data. We provide an example using the case of the COVID-19 outbreak that happened on board of USS Theodore Roosevelt in early 2020.
	\end{abstract}
	
	
	\textbf{Keywords:} Risk, pandemics, epidemiological surveillance, decision support, machine reasoning, Bayesian causal network.

	\maketitle

	\section{Introduction}
	\label{sec:}

	Epidemiological Surveillance (ES) uses various models to forecast the spread of infectious disease in real-time. The ES models can predict the pandemic's mortality, but they do not account for uncertainties such as reliability of testing technology, specific environmental and social factors. In the context of preparedness for future pandemics, they also do not account for ``the availability of treatment, clinical support, and vaccines'' \cite{reed2013novel}. 
	
	Some ES models can be ``stratified'' for age, gender, or other variables, but do not provide any causal analysis of those and the risks of interest that must be assessed by the preparedness decision-makers. In other words, they are not presented in the form that enables its use in pandemics analysis, including sensitivity analysis and model explainability.

\begin{figure}[!ht]
		\begin{center}
			\includegraphics[width=0.49\textwidth]{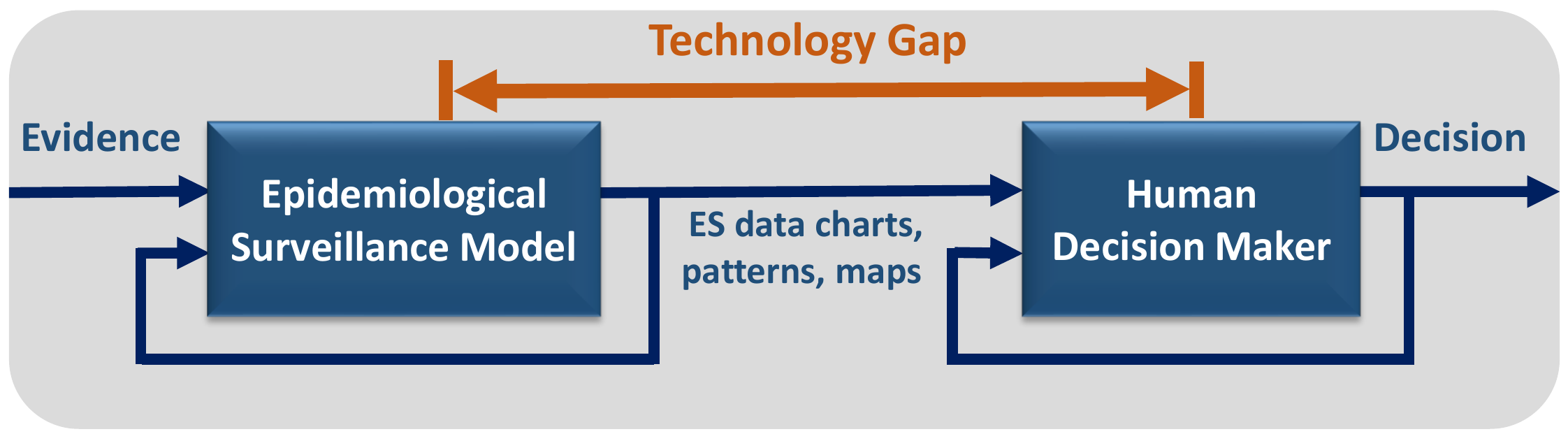}
		\end{center}
		\caption{The technology gap between modern ES and human decision-making: the result of the epidemic/pandemic modeling is provided directly to human experts. This output does not usually include the results of new knowledge inference, or reasoning, to help experts to cope with the abundance of data and uncertainty.}
		\label{fig:Gap}
	\end{figure}

	COVID-19 outbreak provided valuable lessons and unveiled critical disadvantages of the existing ES models including the following:
	\begin{enumerate}
		\item The ES model's outcomes need to be further translated to become usable for human decision-makers. There is a technology gap between the existing models and the decision-making process, as illustrated in Fig. \ref{fig:Gap}.
		\item This can be implemented using computational intelligence (CI) support in the decision-making process.
		\item The CI tool needs to be based on causal models that account for uncertainties, as well as perform fusion and forecasting on those uncertainties. 
	\end{enumerate}

	The answers to the above challenges lie in the usage of machine reasoning, namely, causal models such as causal Bayesian networks. These causal models operate using probabilities, thus accounting for uncertainties, and enable knowledge inference based on priors and evidence \cite{pearl2019seven}. This approach has been applied to risk assessment in multiple areas of engineering and business \cite{fenton2012risk}, risk profiling in identity management \cite{yanushkevich2018understanding,yanushkevich2019cognitive1}, medical diagnostics \cite{vinarti2019personalized}, and very recently to the analysis of COVID-19 risks such as fatality and disease prevalence rates \cite{neil2020bayesian}.
	
	The causal models shall be the core component of the Decision Support System (DSS) that would support human-decision makes in assessing the risk of the infectious disease outbreaks.
	
	The DSS concept was once known as an ``expert system'' that provides certain automation of reasoning (though mainly based on deterministic rules rather than Bayesian approach) and interpretation strategies to extend experts' abilities to apply their strategies efficiently \cite{zachary1997decision}. Examples of contemporary DSS are personal health monitoring systems \cite{andreu2016wize}, e-coaching for health \cite{ochoa2018architecting}, security checkpoints \cite{labati2016biometric,yanushkevich2019cognitive2}, and multi-factor authentication systems \cite{roy2018fuzzy}.

	This paper focuses on developing a DSS for ES, with a CI core based on causal Bayesian networks. We define a DSS as a crucial bridging component to be integrated into the existing ES systems in order to provide situational awareness and help handle the outbreaks better.

	This paper is organized as follows. Contributions are listed in Section \ref{sec:Contr}. Definitions of the relevant concepts are given in Section \ref{sec:Background}. The DSS concept and the fundamental risk assessment operations are described in Section \ref{sec:Cogn}. An example of reasoning on a causal network for the case of the USS Theodore Roosevelt is shown in Section \ref{sec:Exp}. Section \ref{sec:Concl} concludes the paper.

	
	\begin{figure}[!ht]
		\begin{center}
			\includegraphics[width=0.49\textwidth]{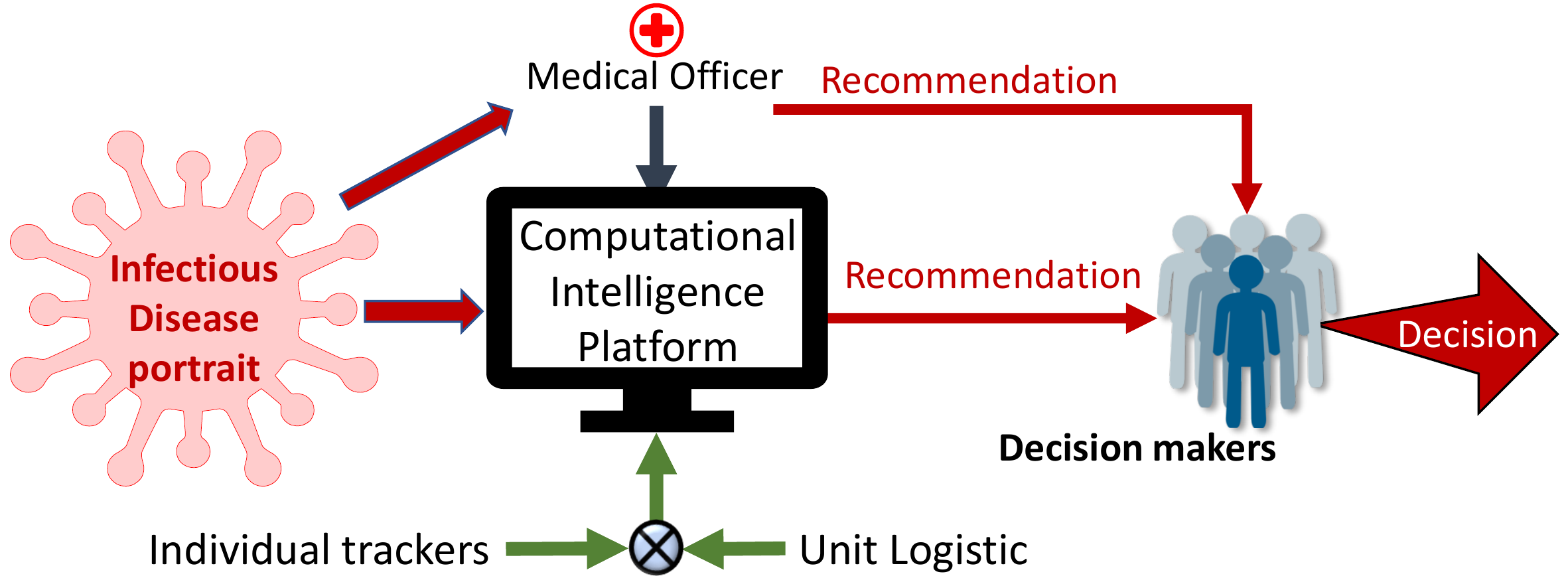}
		\end{center}
		\caption{The overall architecture for the proposed decision support system. We focus on a specific case scenario of COVID-19 on the USS Theodore Roosevelt. Therefore, the team leader, captain of the ship, is given recommendations by both the medical officer and the support system.}
		\label{fig:Support}
	\end{figure}

	\section{Contribution}
	\label{sec:Contr}
	Fig. \ref{fig:Support} illustrates the general framework for the proposed Decision Support System (DSS). In this paper, we illustrate the idea of using machine reasoning to assist the team leaders in making decisions by recommending the best course of action given evidence. Using the USS Theodore Roosevelt as a case study, we explore how different preventive behaviors or measures, such as wearing face masks, impact the chance of subjects being infected with COVID-19. Since all data regarding this case study is obtained after the fact, all reasoning is based on evidence (reactive) and not proactive. It stands to reason that when using these data as well as the fusion of various current and past heterogeneous variables, we can accumulate new knowledge for predicting the impact of future outbreaks, and help prepare for those.
	
	In this study, we \textbf{identify a technological gap} in the ES in both technical and conceptual domains (Fig. \ref{fig:Gap}). Conceptually, the ES users require significant cognitive support using the CI tools. This paper addresses the key research question: \textbf{How to bridge this gap using the DSS concept?} We follow a well-identified trend in academic discussion on the future generation DSS \cite{lai2020assessing}. 
	
	This paper make further steps and contributes to the practice of technology gap bridging. The key contribution is twofold:
	\begin{enumerate}
		\item Development of a \textbf{reasoning and prediction mechanism, the core of a DSS}; for this, a concept of a Bayesian causal network \cite{pearl1988probabilistic} is used; in particular, a recent real-world scenario of COVID-19 was described using a Bayesian network \cite{neil2020bayesian}.
		\item Development of the \textbf{complete spectrum of the risk and bias measures}, including ES taxonomy updating. 
	\end{enumerate}
	These results are coherent with the solutions to the following related problems:
	\begin{enumerate}
		\item [$-$] The technology gap ``pillars'' in Fig. \ref{fig:Gap} are the Protocol of the ES model and the Protocol of the DSS. These protocols are different, e.g. spread virus behavior and conditions of small business operation. The task is to convert the ES protocol into a DSS specification. Criteria of efficiency of conversion are an acceptance of a given field expert. \textbf{The reasoning mechanism based on the causal network intrinsically contains the protocol conversion}. We demonstrate this phenomenon in our experiments.
		\item [$-$] The DSS supports an expert to make decisions under uncertainty in a specific field of expertise. Specifically, intelligent computations help an expert in better interpretation of uncertainty under chosen precautions. \textbf{The risk and bias are used in this paper as a precaution of different kinds of uncertainties} related to ES data \cite{dicker2006principles}, testing tools, human factors, ES model turning parameters, and artificial intelligence. 
	\end{enumerate} 
	A DSS concept suitable for the ES model is proposed in this paper.

	\section{Background}
	\label{sec:Background}
	
	In our study, we model the DSS as a complex multi-state dynamic system \cite{yanushkevich2019cognitive2}. A cognitive DSS is a semi-automated system, which deploys CI to process the data sources and to assess risks and other ``precaution'' measures such as trust in the CI and various biases influencing the decision \cite{yanushkevich2019cognitive1}. The crucial idea of our approach is that the risk assessment should be performed using the reasoning mechanism \cite{pearl2019seven}. This assessment is submitted to a human operator for the final decision.



	\section{Core Support System}\label{sec:Cogn}
	
	The core of the proposed DSS is a causal network that allows us to perform reasoning. The reasoning operations are defined as follows:

	\emph{1) Prior data representations and assessments,} such as statistics and distribution of data after an outbreak that has already happened, as well as the pre-existing conditions. In causal modeling, the priors are represented by a corresponding probability distribution function \cite{pearl2019seven}. 
	
	\emph{2) Causal analysis} is based on the ``cause-effect'' paradigm \cite{pearl1988probabilistic}. Another advanced tool is Granger causality analysis, usually used to analyze time series and to determine whether one can forecast the other \cite{spirtes2016causal}.
	
	
	
	
	\emph{3) Reasoning} is the ability to form an intelligent conclusion or judgment using the evidence. Causal reasoning is a judgment under uncertainty performed on a causal network \cite{pearl1988probabilistic}. 
	
	
	
	\emph{4) Prediction.} In complex systems, meta-learning and meta-analysis are be used to predict the overall success or failure of the predictor. The most valuable information is in the ``tails'' of the probabilistic distributions \cite{davison2015statistics,stehlik2010favorable}. 


	\subsection{Causal network}
	
	A causal network is a directed acyclic graph where each node denotes a unique random variable. A directed edge from node \(X_1\) to node \(X_2\) indicates that the value of \(X_1\) has a direct causal influence on the value of \(X_2\). Uncertainty in causal networks is represented as \emph{Conditional Uncertainty Tables} (CUTs). A CUT is assigned to each node in the causal network, and it is a table that is indexed by all possible value assignments to the parents of this node. Thus, each entry of the CUT is a model of a conditional ``uncertainty'' that varies according to the choice of the uncertainty metric.
	
	%
	
	A recent review \cite{rohmer2020uncertainties} describes the various types of causal networks that are deployed in machine reasoning:
	\begin{itemize}
		\item Bayesian \cite{pearl2019seven},
		\item Imprecise \cite{Coolen2011}, 
		\item Interval \cite{de1994probability},
		\item Credal \cite{cozman2000credal},
		\item DS \cite{simon2008bayesian},
		\item Fuzzy \cite{baldwin2003inference}, 
		\item Subjective \cite{ivanovska2015subjective}.
	\end{itemize}

	The type of a causal network can be chosen given the DSS model and a specific scenario. The choice depends on the CUT as a carrier of \emph{primary} knowledge and as appropriate to the scenario.
	Various causal computational platforms for modeling several systems were compared, in particular, in \cite{misuri2018tackling} (Dempster-Shafer vs. credal networks), and \cite{yanushkevich2018understanding,yanushkevich2019cognitive1} (Bayesian vs. interval vs. Dempster-Shafer vs. fuzzy networks). 

	\subsection{Bayesian causal network}
	
	In our study, we use Bayesian causal networks, often simply called Bayesian networks. Our motivation for choosing this type of causal networks is driven by the fact that the Bayesian (probabilistic) interpretation of uncertainty provides acceptable reliability for decision-making. 
	
	The Bayesian decision-making is based on evaluation of a \emph{prior} probability given a \emph{posterior} probability and \emph{likelihood} (event happening given some history of previous events).
	\begin{center}
		\mbox{
			\begin{footnotesize}
				$\overbrace{P(\texttt{Hypothesis|Data})}^{Prior}=
				\overbrace{P(\texttt{Data|Hypothesis})}^{Likelihood}\times
				\overbrace{P(\texttt{Hypothesis})}^{Posterior}$
			\end{footnotesize}
	}\end{center}
	
	In a Bayesian network, the nodes of a graph represent random variables $X=\{X_1, \ldots, X_m\}$ and the edges between the nodes represent direct causal dependencies. To construct a Bayesian network, factoring techniques are generally applied. Thus, this network is based on a \emph{factored} representation of joint probability distribution:
	\begin{equation}
		P(X)=\overbrace{\prod_{i=1}^{m}P(X_i|\overbrace{\texttt{Par}(X_i)}^{Nodes})}^{Factorization}
	\end{equation}
	where $\texttt{Par}(X_i)$ denotes a set of parent nodes of the random variable $X_i$. The nodes outside $\texttt{Par}(X_i)$ are conditionally independent of $X_i$. 
	
	The posterior probability of $X_1$ is called a \emph{belief} for $X_1$,	$\texttt{Bel}(X_1)$, and the probability $P(X_1|X_2)$ is called the \emph{likelihood} of $X_1$ given $X_2$ and is denoted $L(X_1|X_2)$.
	
	While the Bayesian network structure reflects the causal relationships, its probability reflects the strengths of the relationships.

	\subsection{Risk Assessment}
	
	Risk and other ``precaution'' measures such as bias and trust are often used to evaluate a cognition-related performance in cognitive decision support systems \cite{lai2020assessing}.
	
	The risk and trust measures are used in ES in simple forms such as `high-risk group', `risk factor', and `systematic difference in the enrollment of participants' \cite{dicker2006principles}. However, the cognitive DSS is expected to provide the experts with detailed assessments of epidemic scenarios and make the decision process more transparent and explainable. For example, syndrome surveillance consists of real-time indicators for a disease allowing for early detection \cite{dicker2006principles}. 
	
	The experts seek the DSS support in answering the following questions: 
	\begin{itemize}
		\item What are the risks given the state of the disease outbreak and the health care resources? 
		\item How reliable are the surveyed or collected data? 
		\item What kind of biases are present or expected in data collection, algorithms, and CI decision making?
	\end{itemize}
	
	
	Below, we provide a definition of risk by the US' National Institute of Standards and Technology (NIST) \cite{nist2017security}.
	
	\begin{defn}\label{def:Risk}
		\textbf{Risk} is a measure of the extent to which an entity is threatened by a potential circumstance or event, and typically is a function of: (i) the adverse \textbf{impact}, also called \textbf{cost} or magnitude of harm, that would arise if the circumstance or event occurs, and (ii) the \textbf{likelihood} of event occurrence. 
	\end{defn}
	For example, in automated decision making, and in our study, the \texttt{Risk} is defined as a function $F$ of Impact (also known as Cost), $I$, of a circumstance or event and its occurrence probability, $P$: 
	\begin{equation}\label{eq:risk}
	\texttt{Risk} = F(\texttt{I},\ \texttt{P})
	\end{equation}
	In other words, risk of event represents its impact provided given the likelihood of the decision error. 

	The other ``precaution'' measures often used in causal models include bias and trust to the CI recommendations \cite{lai2020assessing}.

	\begin{defn}\label{def:Bias}
		\textbf{Bias} in the ES refers to the tendency of an assessment process to systematically over- or under-estimate the value of a population parameter.
	\end{defn}
	For example, in the context of detecting or testing for infectious disease, the bias is related to the sampling approaches (e.g. tests are performed on a proportion of cases only), sampling methodology (systemic/random or ad-hoc), and chosen testing procedures or devices \cite{world2015manual}.
	The biases are probabilistic in nature because the evidence and information gathered to make a decision is incomplete, inconclusive, ambiguous, conflicting, and has various degrees of believability. 	
	Identifying and mitigating bias is essential for assessing decision risks and CI biases \cite{whittaker2018ai,gates2018technology,lai2020assessing}. 	
	

	
	Acceptance of the cognitive DSS technology by human decision-makers is determined by the combination of the bias, trust, and risk factors \cite{anand2013pruning,feng2014security}. Other contributing factors include belief, confidence, experience, certainty, reliability, availability, competence, credibility, completeness, and cooperation \cite{zhang2014trust,cho2015survey}. 
	In our approach, the causal inference platform calculates various uncertainty measures \cite{yanushkevich2018understanding} in risk and bias assessment scenarios.
	


	\subsection{Reasoning for infection outbreak and impact prediction}
	
	Probabilistic reasoning on causal (Bayesian) networks enables knowledge inference based on priors and evidence has been applied to diagnostics for precision medicine \cite{vinarti2019personalized}. Recently, COVID-19 risks analysis was performed by \cite{neil2020bayesian}: the Bayesian inference was applied to learn the proportion of the population with or without symptoms from observations of those tested along with observations about testing accuracy.

	\section{The causal model for USS Theodore Roosevelt Ship}\label{sec:Exp}
	During the time of deployment of the USS Theodore Roosevelt Ship around mid-January, an outbreak of COVID-19 occurred that affected marines (younger healthy adults). Approximately 1000 of the 1417 service members were determined to be infected with COVID-19. An investigation during April 20-24, conducted by US Navy and CDC, includes a study on 382 voluntary service members \cite{payne2020sars}.

\begin{figure*}[!ht]
		\centering
		\includegraphics[width=\textwidth]{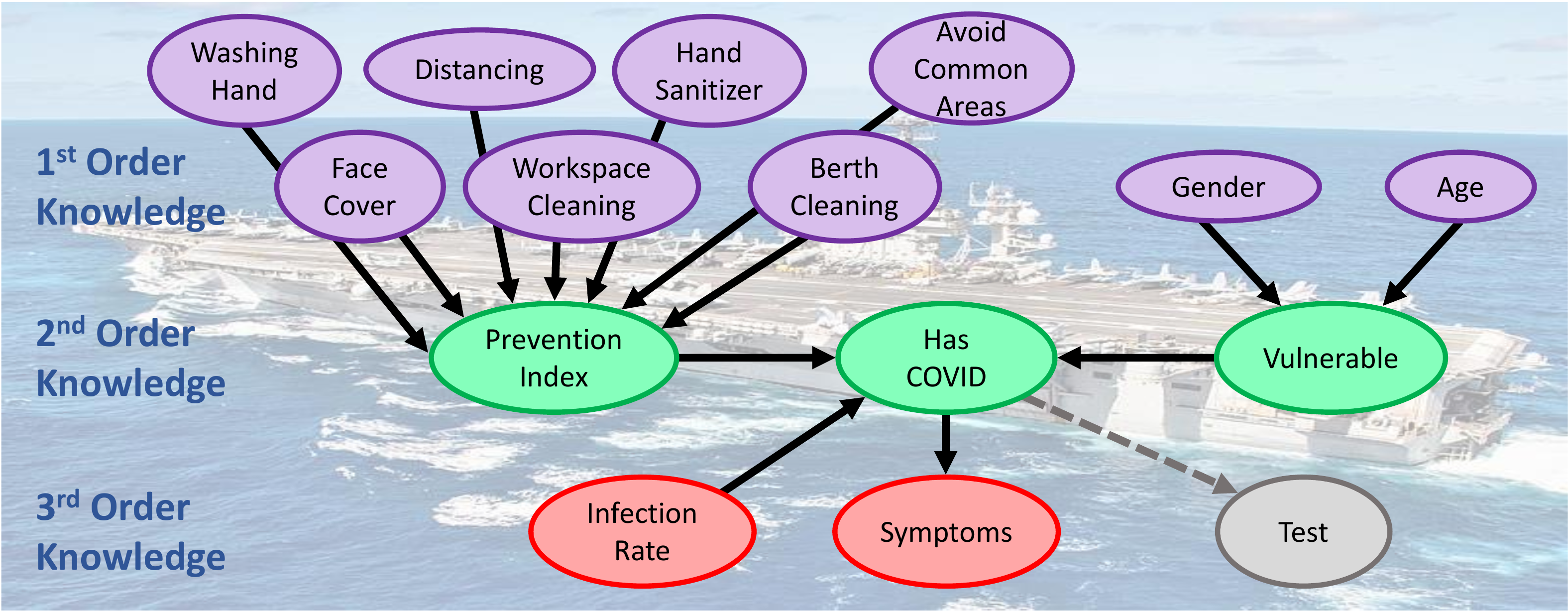}
		\caption{Causal network for the case of the USS Theodore Roosevelt data. 1$^{st}$, 2$^{nd}$, and 3$^{rd}$ order knowledge is represented in purple, green, and red, respectively, while grey-color indicates inactive nodes. The background image source: Wikipedia}
		\label{fig:BN1}
	\end{figure*}

	In our study, we created a fragment of a causal network based on the available data (Figure~\ref{fig:BN1}). The risks assessed in the DSS using the causal network include the `Infection Rate', False Positive Rate (FPR), False Negative Rates (FNR). We define the 1$^{st}$, 2$^{nd}$, and 3$^{rd}$ order knowledge as the prior, calculated, and inferred knowledge, respectively.
	
	\subsection{Data Collection \& Preparation}
	
	The causal network for this example is a Bayesian network (BN), with Conditional Probability Tables (CPTs) assigned to the nodes. The CPTs were constructed using the data retrieved from \cite{payne2020sars}. 
	
	Given the reported test results based on two types, enzyme-linked immunosorbent assay (ELISA) and real-time reverse transcription-polymerase chain reaction (RT-PCR), error rates such as FPR and FNR can be estimated. The designed BN includes a `\texttt{Test} node representing the results of ELISA and the previous RT-PCR test results. Using these results, FNR and FPR are computed as follows:
	\begin{align}\label{eq:rates}
	\text{FPR} = \frac{FP}{FP+TN}, & 
	& 
	\text{FNR} = \frac{FN}{FN+TP}
	\end{align}
	where True Negative ($TN$) represents a healthy subject reported (by testing) as healthy, True Positive ($TP$) represents an infected subject reported (by testing) as infected, False Positive ($FP$) represents a healthy subject reported (by testing) as infected, and False Negative ($FN$) represents an infected subject reported (by testing) as healthy.
	%
	
	In this paper, we measure the `Infection Rate', defined as the ratio between the number of infections and the amount of population at risk:
	\begin{equation}
	\text{Infection Rate}= K \times \frac{\text{number of infections}}{\text{Population}}
	\end{equation}
	where $K$ is a constant value which we set to be 100 in order for the Infection Rate value to be within the interval 0 to 100.
	Table~\ref{tab:nodes} shows all the nodes in the BN and their corresponding states and probabilities. The probabilities for the prior nodes in Table~\ref{tab:nodes} are captured based on the statistics collected in \cite{payne2020sars}. For example, in \cite{payne2020sars} there is a total of 382 volunteers, of which 351 volunteers reported washing their hands as a prevention measure. This results in the probabilities of 8.12\% (31/382) of volunteers who were not washing their hands, and 91.88\% (351/382) of those washing their hands. In this paper, we assume a uniform distribution for the node `\texttt{Infection Rate}' as no value was given in \cite{payne2020sars}. It should be noted that this value is approximately 70\% (1000/1417) for the USS Theodore Roosevelt, based on the reported results \cite{payne2020sars}.
	
	\begin{table}[h]
		\centering
		\caption{Prior nodes and states defined based on data collected.}
		\label{tab:nodes}
		\begin{tabular}{ll}
			\hline
			\hline
			\textbf{Node} & \textbf{States (proportions)} \\[0.2em] 
			\hline
			`\texttt{Hand Wash}' & \emph{no} (8.12\%); \emph{yes} (91.88\%) \\
			`\texttt{Hand Sanitizer}' & \emph{no} (6.81\%); \emph{yes} (93.19\%) \\
			`\texttt{Avoid Common Areas}' & \emph{no} (62.04\%); \emph{yes} (37.96\%) \\
			`\texttt{Face Cover}' & \emph{no} (25.92\%); \emph{yes} (74.08\%) \\
			`\texttt{Workspace Cleaning}' & \emph{no} (19.63\%); \emph{yes} (80.37\%) \\
			`\texttt{Berth Cleaning}' & \emph{no} (34.03\%); \emph{yes} (65.97\%) \\
			`\texttt{Keeping Distance}' & \emph{no} (49.74\%); \emph{yes} (50.26\%) \\
			`\texttt{Infection Rate}' & \emph{0 to 10} (Uniform Distribution) \\
			`\texttt{Gender}' & \emph{male} (75.65\%); \emph{female} (24.35\%) \\
			\multirow{2}{*}{`\texttt{Age}'} & \emph{18-24} (29.58\%); \emph{25-29} (20.42\%) \\
			&\emph{30-39} (38.74\%); \emph{40-59} (11.26\%)\\
			\hline
			\hline
		\end{tabular}
	\end{table}
	
	In this paper, we introduce a measure called the Preventive Index (\texttt{PI}) that illustrates the idea of how selected actions can indirectly increase/decrease the chance of infection. \texttt{PI} for a specific action is defined as follows:
	\begin{equation}\label{eq:pi}
	\texttt{PI}_i= 1+\frac{\beta_{i}-\alpha_{i}}{\beta_{i}}
	\end{equation}
	where subscript $i$ represents one of the prevention measures, $\alpha_{i}$ represents the probability of having COVID-19 given that prevention action $i$ is performed, and $\beta_{i}$ represents the probability of having COVID-19 given no prevention measure $i$ implemented. In other words, it captures the degree of influences of each preventive measure on the infection rate. The probability values, $\alpha$ and $beta$, are calculated based on the statistics from \cite{payne2020sars}, and are summarized in Table \ref{tab:nodes2}.
	
	For example, it was reported that 283 volunteers used face covers as a preventive measure, of which 158 reported having COVID-19. In addition, it is known that a total of 238 volunteers were having COVID-19. Therefore, combining both knowledges, we get 55.83\% (158/283) and 80.81\% (80/99) of having COVID-19 for the subjects who used face cover or not, respectively. Based on Equation (\ref{eq:pi}), a \texttt{PI} of 1.3091 ($1+(80.81-55.83)/80.81$) is obtained for using a face cover. This represents a ``positive'' \texttt{PI}, therefore, reducing the overall probability of infection.
	
	The CPT for `\texttt{Prevention Index}' represents a distribution of the cumulative prevention index. It is calculated based on the product of the individual preventive indices:
	\begin{equation}\label{eq:cumu}
	\texttt{PI}_{cumulative}=\texttt{PI}_{0}\prod_{i=1}^{N}\gamma_i\texttt{PI}_{i}+(1-\gamma_i)
	\end{equation}
	where $N$ is the number of prevention measures (7 in this paper), $\gamma_i$ is a binary value indicating whether the prevention measure $i$ is taken, $\texttt{PI}_{i}$ is the individual prevention index for behavior/action $i$, and $\texttt{PI}_{0}$ represents the default prevention index for no preventive measure taken (it is assumed to be 1 in this paper).

	\begin{table}[h]
		\centering
		\caption{Chance of having COVID-19 given the prevention measures.}
		\label{tab:nodes2}
		\begin{tabular}{lccc}
			\hline
			\hline
			&	Doing 	&	Not Doing 	&	Prevention 	\\
			& ($\alpha$) & ($\beta$) & Index\\
			\hline
			`\texttt{Hand Wash}' 	&	62.11	&	64.52	&	1.0373	\\
			`\texttt{Hand Sanitizer}' 	&	61.52	&	73.08	&	1.1582	\\
			`\texttt{Avoid Common Areas}' 	&	53.79	&	67.51	&	1.2032	\\
			`\texttt{Face Cover}' 	&	55.83	&	80.81	&	1.3091	\\
			`\texttt{Workspace Cleaning}' 	&	63.52	&	57.33	&	0.8921	\\
			`\texttt{Berth Cleaning}' 	&	61.90	&	63.08	&	1.0186	\\
			`\texttt{Keeping Distance}' 	&	54.69	&	70.00	&	1.2188	\\
			\hline
			\hline
		\end{tabular}
	\end{table}
	
	For example, a base case where no preventive action is used will result in a prevention index of 1. Based on Equation (\ref{eq:cumu}), if `\texttt{Hand Wash}' is applied, the index is increased to 1.0373 ($1\times1.0373$). This can be further increased to 1.2014 ($1\times1.0373\times1.1582$) if `\texttt{Hand Sanitizer}' is used with `\texttt{Hand Wash}'. Note that a special case exists for the `\texttt{Workspace Cleaning}' action as it reduces the index instead of increasing. Given this definition, the prevention index ranges from 0.8921 (only `\texttt{Workspace Cleaning}') to 2.3492 (all actions except `\texttt{Workspace Cleaning}') depending on the number of preventive action taken.
	
	Similarly, the computation of the CPTs for `\texttt{Vulnerable}' (\texttt{V}) uses the normalized values of the product of the probabilities for `\texttt{Gender}' (\texttt{G}) and `\texttt{Age}' (\texttt{A}):
	\begin{equation}\label{eq:vul}
	V=\frac{A_i\times G_j}{(A_i\times G_j) + ((1-A_i)\times (1-G_j))}
	\end{equation}
	where $A_i$ represents the probability of having COVID-19 for age category $i$ (18-24, 25-29, 30-39, and 40-59), and $G_i$ represents the probability of having COVID-19 for gender category $j$ (male or female).
	
	For example, given that the probability of having COVID-19 for the age group 18-24 is 68.1\% and the probability of having COVID-19 for gender group male is 65.7\%, the degree of vulnerability is computed as follows (Equation (\ref{eq:vul})):
	\begin{equation}
	V= \frac{0.681\times0.657}{0.681\times0.657 + 0.319\times0.343} = \fbox{0.8035} \nonumber
	\end{equation}

	As a sample, two CPT tables, for the Nodes ``Symptoms'' and ``Vulnerable'' are shown in Tables \ref{tab:CPTs} and \ref{tab:CPTv}, respectively.
	
	\begin{table}[h]
		\centering
		\caption{CPT for Symptoms} \label{tab:CPTs}
		\begin{tabular}{cccccc}
			\hline
			\hline
			\multirow{2}{*}{has COVID}	&	\multicolumn{5}{c}{Symptoms}\\
			&	0	&	1-3	&	4-5	&	6-8	&	$>$8	\\
			\hline
			No	&	\cca{0.3750}	&	\cca{0.3403}	&	\cca{0.0903}	&	\cca{0.1111}	&	\cca{0.0833}	\\
			Yes	&	\cca{0.1849}	&	\cca{0.2143}	&	\cca{0.1555}	&	\cca{0.2101}	&	\cca{0.2353}	\\
			\hline
			\hline
		\end{tabular}
	\end{table}

	\begin{table}[h]
		\centering
		\caption{CPT for Vulnerable} \label{tab:CPTv}
		\begin{tabular}{cccc}
			\hline
			\hline
			\multirow{2}{*}{Gender}	&	\multirow{2}{*}{Age}	&	\multicolumn{2}{c}{Vulnerable}			\\
			&		&	No	&	Yes	\\
			\hline
			\multirow{4}{*}{Male}	&	18-24	&	\cca{0.1965}	&	\cca{0.8035}	\\
			&	25-29	&	\cca{0.2262}	&	\cca{0.7738}	\\
			&	30-39	&	\cca{0.2678}	&	\cca{0.7322}	\\
			&	40-59	&	\cca{0.2926}	&	\cca{0.7074}	\\
			\multirow{4}{*}{Female}	&	18-24	&	\cca{0.3053}	&	\cca{0.6947}	\\
			&	25-29	&	\cca{0.3444}	&	\cca{0.6556}	\\
			&	30-39	&	\cca{0.3966}	&	\cca{0.6034}	\\
			&	40-59	&	\cca{0.4263}	&	\cca{0.5737}	\\
			\hline
			\hline
		\end{tabular}
	\end{table}

	The CPT for `\texttt{Has COVID}' is estimated based on the relationship of `\texttt{Prevention Index}' (\texttt{PI}), `\texttt{Vulnerable}' (\texttt{V}) and `\texttt{Infection Rate}' (\texttt{IR}), specifically:
	\begin{equation}\label{eq:covid}
	P(\texttt{Has COVID})=\frac{\texttt{IR}}{\texttt{PI}}\times(\texttt{V}+1)
	\end{equation}
	In Equation \ref{eq:covid}, the \texttt{IR} is reduced based on the \texttt{PI} and then multiplied by the value $\texttt{V}+1$. This serves as a multiplier when calculating the chance of having COVID-19. The value $V=0$ (\texttt{False}) means that the person is not vulnerable. In this paper, we assume $V=1$ (\texttt{True}) when a subject is vulnerable. Thus, the vulnerable subject's chance of having COVID-19 is multiplied by a factor of $V+1=2$.
	
	The remaining nodes `\texttt{Symptoms}' and `\texttt{Test}' have CPTs created directly based on the data from \cite{payne2020sars}. For example, the conditional probability for 0 symptoms is defined as $P(\texttt{Symptoms}=0|\texttt{COVID}=Yes)=11.52\%$ and $P(\texttt{Symptoms}=0|\texttt{COVID}=No)=14.14\%$.
	
	The list of the COVID-19 symptoms reported in \cite{payne2020sars} included loss of taste, smell, or both, palpitations, fever, chills, myalgia, cough, nausea, fatigue, shortness of breath/difficult breathing, chest pain, abdominal pain, runny nose, diarrhea, headache, vomiting, and sore throat. Note that in this paper, we are interested in the number of symptoms and not their type.
	
	One of our key interest in this paper is to evaluate the risk of being infected with COVID-19. Equation (\ref{eq:risk}) defines risk as a function of impact and probability of the event of interest. In this paper, we define two types of infection risk: the risk of missing the true infection (positive risk, $\text{Risk}_{\texttt{p}}$) and the risk of declaring the infection when it is not the case (negative risk, $\text{Risk}_{\texttt{n}}$). A positive risk reflects the impact of virus spreading undetected, while a negative risk is determined by the false positive testing resulting in unnecessary treatment or quarantine. These risk equations are defined as follows:
	\begin{eqnarray}
	\text{Risk}_{\texttt{p}} = \texttt{Impact}_{u}\times \text{FNR} +
	\texttt{Impact}_{k}\times \text{TPR} \\
	\text{Risk}_{\texttt{n}}= \texttt{Impact}_{q}\times \text{FPR} + 
	\texttt{Impact}_{c}\times \text{TNR}
	\end{eqnarray}
	where $\texttt{Impact}_u$ represents the cost of undetected virus spreading (very high), $\texttt{Impact}_k$ represents the cost of unnecessary ``precaution'' such as quarantining (high), $\texttt{Impact}_q$ represents cost of quarantine (low), $\texttt{Impact}_c$ represents the base cost of testing (very low), True Positive Rate (TPR) is defined as $\text{TPR} = 1-\text{FNR}$, True Negative Rate (TNR) is defined as $\text{TNR} = 1-\text{FPR}$ and FPR/FNR is defined by Equation (\ref{eq:rates}).
	
	For example, given a specific scenario where the error rates (FPR = 1\% and FNR = 20\%) and impact values (very high = 4, high = 3, low = 2, very low = 1) are given, the overall risk is estimated as follows:
	\begin{align}
	\text{Risk}_{p}=4\times0.20+3\times0.80=\fbox{3.20} \nonumber\\
	\text{Risk}_{n}=2\times0.01+1\times0.99=\fbox{1.01} \nonumber
	\end{align}
	
	\subsection{Experiments}
	
	Experiments on the BN shown in this section were implemented using the open-source Python library pyAgrum~\cite{gonzales2017agrum}. 
	
	The following scenarios were considered in our experiments:
	\begin{itemize}
		\item Scenario 1: The effects of increasing the `\texttt{Prevention Index}' with a constant `\texttt{Vulnerability}' and `\texttt{Infection Rate}' on `\texttt{Has COVID}';
		\item Scenario 2: The influence of `\texttt{Symptoms}' on `\texttt{Has COVID}';
		\item Scenario 3: The impact of `\texttt{Has COVID}' on `\texttt{Vulnerable}'; and
		\item Scenario 4: The predicted `\texttt{Infection Rate}' given `\texttt{Symptoms}' and `\texttt{Prevention Index}'
	\end{itemize}

	\subsubsection*{Scenario 1}
	Scenario 1 explores how different preventive measures influence the chance for a subject to be infected with COVID-19. Table \ref{tab:Scenario1} shows the probability of a subject having COVID-19 given various `\texttt{Prevention Index}' and whether or not they are `\texttt{Vulnerable}'. As the prevention index increases, the chance of getting COVID-19 decreases regardless of vulnerability. For example, when (a) prevention index 0.9 and 1.0, the chance of getting COVID-19 is reduced from 82.23\% to 79.41\%, which represents about 2.82\% difference in getting COVID-19 for a 0.1 increase in prevention index. Table \ref{tab:Scenario1} (c) assumes a specific case where no evidence for `\texttt{Vulnerable}' is given.
	
	\begin{table*}[h]
		\centering
		\caption{Scenario 1: Effects of `\texttt{Prevention Index}' on `\texttt{Has COVID}' with $\texttt{Infection Rate}=70\%$}
		\label{tab:Scenario1}
		\begin{tabular}{ccc}
			(a) $\texttt{Vulnerable}=Yes$ & (b) $\texttt{Vulnerable}=No$ & (c) $\texttt{Vulnerable}=?$ \\
			\begin{tabular}{cc}
				\hline
				\hline
				Prevention & \multirow{2}{*}{$P(\texttt{Has COVID}=\texttt{Yes})$}	\\
				Index &	\\
				\hline
				0.9	&		82.23	\\
				1.0	&		79.41	\\
				1.5	&		67.78	\\
				2.0	&		59.12	\\
				2.3	&		54.94	\\
				No Evidence		&	67.09	\\
				\hline
				\hline
			\end{tabular}&
			\begin{tabular}{cc}
				\hline
				\hline
				Prevention & \multirow{2}{*}{$P(\texttt{Has COVID}=\texttt{Yes})$}	\\
				Index &	\\
				\hline
				0.9	&	49.09	\\
				1.0	&	47.09	\\
				1.5	&	39.13	\\
				2.0	&	33.47	\\
				2.3	&	30.79	\\
				No Evidence	&	38.67	\\
				\hline
				\hline
			\end{tabular}&
			\begin{tabular}{cc}
				\hline
				\hline
				Prevention & \multirow{2}{*}{$P(\texttt{Has COVID}=\texttt{Yes})$}	\\
				Index &	\\
				\hline
				0.9	&	74.38	\\
				1.0	&	71.71	\\
				1.5	&	60.81	\\
				2.0	&	52.79	\\
				2.3	&	48.92	\\
				No Evidence	&	60.17	\\
				\hline
				\hline
			\end{tabular}\\
		\end{tabular}
	\end{table*}
	
	\subsubsection*{Scenario 2}
	For scenario 2, we analyze the causal relationship between the number of symptoms and the fact of having COVID-19. Table \ref{tab:Scenario2} shows the chance of having COVID-19 given the number of symptoms. As the number of symptoms increases, the chance of the individual having COVID-19 increases proportionally. For 0 symptoms (a subject is asymptomatic), a chance of having COVID-19 is 29.67\%. It increases to 59.57\% for the subjects having 4-5 symptoms. 
	
	\begin{table}[h]
		\centering
		\caption{Scenario 2: Influences of `\texttt{Symptoms}' on `\texttt{Has COVID}'}
		\label{tab:Scenario2}
		\begin{tabular}{cc}
			\hline
			\hline
			Symptoms	&	$P(\texttt{Has COVID})$	\\
			\hline
			0	&	29.67	\\
			1-3	&	35.02	\\
			4-5	&	59.57	\\
			6-8	&	61.80	\\
			$>$8	&	70.73	\\
			No Evidence	&	46.11	\\
			\hline
			\hline
		\end{tabular}
	\end{table}
	
	\subsubsection*{Scenario 3}
	Scenario 3 considers how to infer the chance of a subject being vulnerable given the evidence of having COVID-19. Table \ref{tab:Scenario3} illustrates this scenario for determining the vulnerability of a person given their infection status. For instance, when a person is diagnosed with COVID-19, he/she very likely belongs to the group of vulnerable.
	\begin{table}[h]
		\centering
		\caption{Scenario 3: Impacts of `\texttt{Has COVID}' on `\texttt{Vulnerable}'}
		\label{tab:Scenario3}
		\begin{tabular}{cc}
			\hline
			\hline
			Has COVID	&	$P(\texttt{Vulnerable}=\texttt{Yes})$	\\
			\hline
			Yes	&	84.36	\\
			No	&	67.04	\\
			No Evidence	&	75.03	\\
			\hline
			\hline
		\end{tabular}
	\end{table}
	
	\subsubsection*{Scenario 4}
	Scenario 4 estimates the infection rate given various evidence for both the number of symptoms and preventive behaviors. Table \ref{tab:Scenario4} shows the estimated average `\texttt{Infection Rate}' for fixed prevention index of 0.9 (a), 1.5 (b), and 2.3 (c). We observed, quite surprisingly, that the infection rate is minimally impacted by the prevention index. In part, this is because we assumed that `\texttt{Prevention Index}' and `\texttt{Infection Rate}' are independent. The number of symptoms, on the other side, has a direct influence on the estimated infection rate.
	\begin{table*}[h]
		\centering
		\caption{Scenario 4: Estimated `\texttt{Infection Rate}' given `\texttt{Symptoms}' and `\texttt{Prevention Index}'}
		\label{tab:Scenario4}
		\begin{tabular}{ccc}
			(a) $\texttt{Prevention Index} = 0.9 $& (b) $\texttt{Prevention Index} = 1.5 $ & (c) $\texttt{Prevention Index} = 2.3 $ \\
			\begin{tabular}{cc}
				\hline
				\hline
				\multirow{2}{*}{Symptoms}	& Average	\\
				&	$\texttt{Infection Rate}~(\%)$	\\
				\hline
				0	&	48.45	\\
				1-3	&	50.31	\\
				4-5	&	57.72	\\
				6-8	&	58.31	\\
				$>$8	&	60.57	\\
				No Evidence	&	53.46	\\
				
				\hline
				\hline
			\end{tabular}
			&
			\begin{tabular}{cc}
				\hline
				\hline
				\multirow{2}{*}{Symptoms}	& Average	\\
				&	$\texttt{Infection Rate}~(\%)$	\\
				\hline
				0	&	48.84	\\
				1-3	&	50.24	\\
				4-5	&	56.63	\\
				6-8	&	57.20	\\
				$>$8	&	59.51	\\
				No Evidence	&	52.78	\\
				
				\hline
				\hline
			\end{tabular}
			&
			\begin{tabular}{cc}
				\hline
				\hline
				\multirow{2}{*}{Symptoms}	& Average	\\
				&	$\texttt{Infection Rate}~(\%)$	\\
				\hline
				0	&	49.13	\\
				1-3	&	50.18	\\
				4-5	&	55.54	\\
				6-8	&	56.08	\\
				$>$8	&	58.23	\\
				No Evidence	&	52.20	\\
				
				\hline
				\hline
			\end{tabular}
			\\
		\end{tabular}
	\end{table*}
	
	Assuming the specific scenario of USS Theodore Roosevelt with the estimated error rates are estimated assuming that RT-PCR test results are ground truth and ELISA test results are predicted cases. Based on this assumption, FPR and FNR are calculated to be = 10.88\% (16/147) and 9.79\% (23/235), respectively. With the given impact values (very high = 4, high = 3, low = 2, very low = 1), the overall risk is computed as follows:
	\begin{align}
	\text{Risk}_{p}=4\times0.0979+3\times0.9021=\fbox{3.0979} \nonumber\\
	\text{Risk}_{n}=2\times0.1088+1\times0.8912=\fbox{1.1088} \nonumber
	\end{align}
	The maximum value of positive risk is 4 (FNR=100\%) and the minimum value is 3 (FNR=0\%). These two cases represent the extreme cases where either all infected subjects are correctly diagnosed or all of them are misdiagnosed. Similarly, the maximum value of negative risk is 2 (FPR=100\%) and the minimum value is 1 (FPR=0\%). The two cases represent the extreme cases where either all healthy subjects are misdiagnosed or cleared.
	
	In this section, we illustrate the idea of how a decision-maker such as a medical officer and/or a captain, can use the proposed causal network to infer the risks in relation to the actions/decisions. Specifically, we show that inaction (no prevention behavior, $\texttt{Preventive Index}=1$) and infection rate of 70\% results in 71.71\% of the crew being infected. If all beneficial prevention behavior is taken ($\texttt{Preventive Index}=2.3$), the chance for the crew to be infected is reduced to 48.92\%.
	
	\subsection{Limitations}
	Given the proposed causal model, there are several limitation/assumptions including:
	\begin{itemize}
		\item Insufficient amount of data to capture the true causal relationship,
		\item The CPT of the BN nodes are populated based on simplified equations and approximations/assumptions, and
		\item There is also an assumption that each prevention measure is independent, while in reality they might be related.
	\end{itemize}
	
	As indicated earlier, some causal relationships are inferred based on the data. Therefore, in case of insufficient data, some relationships can be misleading and/or missing. For example, data regarding spatial location and congestion of the crew is currently missing. Selected subjects may be required to travel through the ship to the targeted areas due to their duty, and this required movement may result in an increased chance of infection.
	
	As well, the nodes such as `\texttt{has COVID}', `\texttt{Vulnerable}', and `\texttt{Prevention Index}', are populated based on the proposed equations. These equations only illustrate the general, not necessarily the exact relationship, and were derived for the given scenario, and may require modification when transferred to another study. For example, age, gender, and preventive behaviors can greatly increase or decrease the chance of infection, but this relationship cannot be captured by deterministic equations.
	
	In this study, we assume that the seven preventive behaviors reported by the volunteers in \cite{payne2020sars} are independent. This assumption is not sufficient, as subjects can generally be classified as risk-averse or risk-tolerant. Risk-averse individuals are much more likely to take preventive measures, that is, the subject that uses face masks are also the ones who keep a social distance.
	
	In addition, the bias in the sampling of data can severely impact the causal network model, specifically the creation of the CPTs. In the USS Theodore Roosevelt case, there is a significant bias regarding the crew composition of younger males. Based on the collected data \cite{payne2020sars}, age group 18-24 contains the most members but also contributes to the largest percentage of having COVID-19 (68.1\%), whereas the age group 40-59 contains the least amount of infected (55.8\%). This contradicts the belief that older people are more vulnerable. Possible reasons for this contradiction may be that younger people are prone to more interactions, while older people take more precautionary measures, as well as they are are more likely to be of higher rank on the ship and have different duties requiring less contact. Lastly, all the data collected in \cite{payne2020sars} were collected on a volunteer basis, and, therefore, represent only a fraction of the ship population.

	\section{Conclusions and Future Work}\label{sec:Concl}

	This paper contributes to bridging the technology gap (Figure \ref{fig:Gap}) that exists between the contemporary ES models and human expert's limitations to handle uncertainty provided by the model while striving to make reliable decisions. It asserts that the solution lies in deploying the causal networks that capture an \textbf{approximation} of joint probabilistic distributions of epidemiological factors.
	
	We propose a general DSS model with an embedded reasoning mechanism using a causal Bayesian network. This reasoning results in the probabilities and risk assessment of the outcomes of interest, thus providing recommendations to the human decision-makers.

	The DSS ability to support human experts with or without technical background should be estimated using various measures, including the generally used ``technological'' performance measures. The recent emergence of ``precaution'' measures such as risk, trust, and bias address this trend. In this paper, we focus on risk assessment. It should be noted that 
	
	Other \textbf{open applied problems} to be further addressed include studies of other precautionary measures such as bias and trust. These shall reflect various decision-making dimensions:
	\begin{itemize}
		\item [$-$] Technical, e.g. prediction accuracy and throughput \cite{roy2018fuzzy},
		\item [$-$] Social, e.g. trust in CI \cite{whittaker2018ai,danks2017algorithmic},
		\item [$-$] Psychological, e.g., efficiency of human-machine interactions \cite{hu2018computational,montibeller2015cognitive,{hou2010optimizing}}, and
		\item [$-$] Privacy and security domain, e.g., vulnerability of personal data \cite{andreou2017identity,bellovin2019privacy,focus2019data}. 
		\end{itemize}
		
		Finally, in the context of epidemic or pandemic preparedness, the human decision makes may need support as the situation develops (proactive reasoning). Given data from the epidemiological model, the output of such DSS is a result of \emph{dynamic evidential reasoning}. This approach shall be further developed for better managing future epidemics and pandemics.

		\section*{Acknowledgments}
		\begin{small}
			This research was partially supported by the Natural Sciences and Engineering Research Council of Canada (NSERC) through grant ``Biometric-enabled Identity management and Risk Assessment for Smart Cities''. The authors acknowledge Dr. V. Shmerko for valuable ideas and suggestions, and Ivan Hu for helping to collect data on the COVID-19 outbreak case on the USS Theodore Roosevelt."
		\end{small}
		
%

	\end{document}